\documentclass[conference]{IEEEtran}

\usepackage{graphicx} 
\usepackage{booktabs} 
\usepackage{amsmath}
\usepackage{amsfonts}
\usepackage{amssymb}
\usepackage{epstopdf}
\newcounter{MYtempeqncnt}
\usepackage{dblfloatfix}
\usepackage[caption=false,font=footnotesize]{subfig}
\usepackage{color}
\usepackage{cite}
\usepackage{algorithmic}
\usepackage{multirow}
\usepackage{array}

\makeatletter
\g@addto@macro \normalsize {%
 \setlength\abovedisplayskip{5pt}%
 \setlength\belowdisplayskip{5pt}%
}
 \newcommand{\thickhline}{%
    \noalign {\ifnum 0=`}\fi \hrule height 1pt
    \futurelet \reserved@a \@xhline
		}
\makeatother

\begin{document}

\title{Power Talk for Multibus DC MicroGrids: \\ Creating and Optimizing Communication Channels}

\author{\IEEEauthorblockN{Marko Angjelichinoski, \v Cedomir Stefanovi\' c, Petar Popovski}
\IEEEauthorblockA{Department of Electronic Systems, Aalborg University\\
Email: \{maa,cs,petarp\}@es.aau.dk}
}

\maketitle

\begin{abstract}
We study a communication framework for nonlinear multibus DC MicroGrids based on a deliberate modification of the parameters of the primary control and termed \textit{power talk}.
We assess the case in which the information is modulated in the deviations of \emph{reference voltages} of the primary control loops and show that the outputs of the power talk communication channels can be approximated through linear combinations of the respective inputs.
We show that the coefficients of the linear combinations, representing equivalent channel gains, depend on the \emph{virtual resistances} of {the primary control loops}, implying that they can be modified such that effective received signal-to-noise ratio (SNR) is increased.
On the other hand, we investigate {the constraints that power talk incurs on the supplied power deviations.
We show that these constraints translate into constraints on the reference voltages and virtual resistances that are imposed on all units in the system.}
In this regard, we develop an optimization approach to find the set of controllable virtual resistances that maximize SNR under the constraints on the supplied power deviations.
\end{abstract}

\IEEEpeerreviewmaketitle

\section{Introduction}\label{sec:Intro}

MicroGrids (MGs) are localized clusters of small-scale Distributed Generators (DGs) that cover small geographical areas and operate either connected to the main grid or in standalone mode \cite{ref:1}.
The MG control plane is divided into hierarchy of three levels, comprising the primary, secondary and tertiary control levels \cite{ref:2}.
The \emph{primary} control enables the basic operation of the system by regulating the electrical parameters (bus voltage and/or frequency) and keeping the supply-demand power balance to guarantee local stability.
It is implemented in a decentralized manner using the \textit{droop control law} \cite{ref:3,ref:3}, relying only on the local measurements of the controllers.
The upper, \emph{secondary} and \emph{tertiary} control levels optimize the performance of the MG in terms of maximizing the quality of the delivered power under minimal operation cost, and, in order to operate properly, require exchange of local information among the controllers.
Recent approaches suggest to avoid use of external communication systems for MG control applications, due to related costs, complexity and reliability issues \cite{ref:2,ref:3}; rather, the existing power line equipment is also used for communication \cite{ref:4,ref:5,ref:6}.

In \cite{ref:7} we introduced \textit{power talk} - a communication technique over the power lines, developed for direct current (DC) MGs and proposed as an alternative to using external communications for upper layer control.
Power talk is an in-band solution that modulates the information into controlled deviations of the parameters of the primary control loops of the DGs.
In this way, a communication channel is induced over the DC bus voltage level through which information-carrying deviations of the voltage (or, equivalently, power) are disseminated throughout the system, and received and processed by other DG units.
The control frequency of the primary droop controller is typically between $10-1000\,\text{Hz}$, which implies that power talk is a narrowband powerline communication (PLC) solution.
It exhibits conceptual similarities with other existing low-rate PLC standards for communication in the AC distribution grids, such as Ripple Carrier, TWACS and Turtle \cite{ref:61}, which also rely on disturbing the (sinusoidal) voltage wave to transmit information.
However, in contrast to these solutions, power talk requires no additional hardware to generate and process the information signals, as it is implemented in the local primary control loop of the power electronic converters that connect the DGs to the DC buses.
Thus, power talk exhibits the self-sustainability feature of the MG paradigm, drawing its reliability from the reliability of the MG system itself.

The power talk communication channel shows some challenging properties that are not commonly encountered in communication systems:
i) non-linear input-output relation,
ii) configurations of primary controllers of all communicating units jointly determine the values of the observed channel outputs, i.e., voltage/power levels of MG buses, 
iii) configurations of primary controllers are also jointly subject to constraints in terms of the allowed supplied power deviations, and
iv) dependence of the channel outputs on the configuration of the rest of the system, i.e., distribution line impedances and the instantaneous values of the loads that change randomly.
The previous works \cite{ref:7,ref:8,ref:hpl,ref:9,ref:mdpi} investigated the performance of power talk in presence of random load changes, when standard communication techniques, such as line-coding and pilot-sequence based training are applied. 
In addition, \cite{ref:7,ref:8,ref:hpl,ref:9,ref:mdpi} focused on a simple single bus DC MG system where the effect of the distribution lines can be ignored and all units observe the same output.
In this paper we expand the analysis of power talk in several ways, as elaborated below.

\subsection*{Motivation and Contributions}

We assume that a general DC MG with $N$ buses, where each bus hosts a single droop controlled DG unit operated as a voltage source, see Fig.~\ref{BusArch} (more details on DG operation are provided in Section~\ref{sec:Preliminaries}).
Denote the DC voltage level of bus $n$ with $v_n$ and the droop control parameters of the units $x_n,\;r_{n}$, $n=1,...,N$, where $x_n$ is the reference voltage and $r_{n}$ is the virtual resistance.
We consider a variant of power talk in which the signaling is done only through the reference voltage $x_n$ by modulating it around its \emph{nominal} value $x_n^\texttt{n}$:
\begin{equation}\label{new1}
	x_n = x_n^{\texttt{n}} + \Delta x_n, \; n=1,...,N,
\end{equation}
i.e., $\Delta x_n$, $n=1,...,N$, are power talk inputs.
The power talk outputs are the deviations of the bus voltages {$v_n$} around their nominal values $v_n^{\texttt{n}}$:
\begin{equation}
	v_n = v_n^{\texttt{n}} + \Delta v_n, \; n=1,...,N,
\end{equation}
where $v_n = v_n( \Delta x_1,...,\Delta x_N)$ is a non-linear function of the inputs.
This deviations translate into deviations of the output powers $p_n$ around their nominal values $p_n^{\texttt{n}}$:
\begin{equation}
	p_n = p_n^{\texttt{n}} + \Delta p_n, \; n=1,...,N,
\end{equation}
where $p_n =\ p_n(\Delta x_1,...,\Delta x_N)$ is also a non-linear function of the inputs.
Further, in order to maintain the quality of the supplied power, which is one of the main goals of the MG operation, the variances of $\Delta p_n$ should be bounded:
\begin{equation}\label{new2}
	\mathbb{E}[{\Delta p_n^2}] \leq \pi_n^2, \; n=1,...,N, 
\end{equation}
where $\pi_n$ is the power deviation budget of unit $n$.

Under the assumption that the reference voltage deviations are small compared to their nominal values, in this paper we derive the linearized models:
\begin{align}\label{eq:intro3}
\Delta v_n & =\sum_{m=1}^N \breve{h}_{n,m}\Delta x_m, \; n = 1,...,N, \\\label{eq:intro5}
\mathbb{E}[\Delta p_n^2] & =\sum_{m=1}^N\breve{\phi}_{n,m}^2\mathbb{E}[\Delta x_m^2], n = 1,...,N,
\end{align}
where $\breve{h}_{n,m} = \breve{h}_{n,m} (r_{1},...,r_{N}) $ and $\breve{\phi}_{n,m} = \breve{\phi}_{n,m} (r_{1},...,r_{N})$, i.e., they are functions of virtual resistances, as shown in Sections~\ref{sec:DiscreteTimeLinearModel} and \ref{sec:CommunicationArchitetures}.
In other words, the resulting linearized model is such that the units can \textit{control} the channel gains $\breve{h}_{n,m}$ through the values of $r_{n}$, $n,m=1,...,N$.
Moreover, using \eqref{new2} and \eqref{eq:intro5}, we show that constraints on the individual signals $\Delta x_n$ are obtained as solutions of a linear system of inequalities, jointly imposed on droop parameters of all units.
In Section \ref{sec:CaseStudy} we show how to exploit the above properties to optimize the received SNR of the observations of the channel outputs.
Specifically, we show that, when the output $\Delta v_n$ is affected by Gaussian noise, the received SNR can be maximized by optimizing the virtual resistances $r_{n}$ under power deviation budgets $\pi_n$, $n = 1,...,N$.
Finally, in Section~\ref{sec:SNRMaximization}, we present an algorithm to obtain the optimal values of the virtual resistances.
We conclude by noting that the power talk schemes assessed in \cite{ref:7,ref:8} can be derived as special instances of the communication framework developed in this paper.

\section{System Model}\label{sec:SystemModel}

\subsection{Signal Model}\label{sec:Preliminaries}

\begin{figure}
\centering
\includegraphics[scale=0.28]{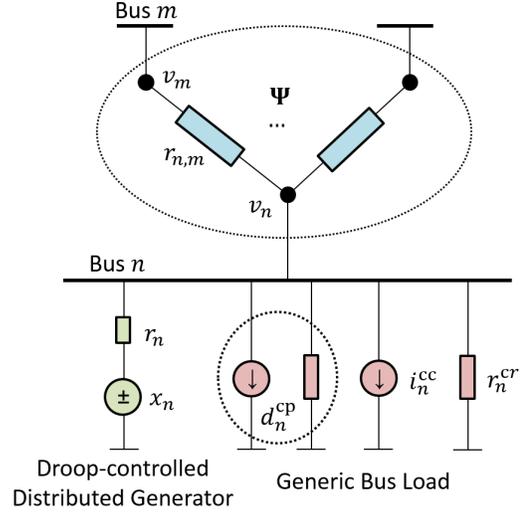}
\caption{General architecture of multibus DC MicroGrid.}
\label{BusArch}
\end{figure}

Fig.~\ref{BusArch} depicts the architecture of a bus in a DC MG system with $N$ buses, where $v_n$ denotes the steady state bus voltage.
Without loss of generality, we assume that each bus hosts a single DG, connected to the distribution infrastructure through a power electronic converter.
On the primary control level, the converter is configured as a voltage source converter (VSC) \cite{ref:3,ref:6}, regulating the bus voltage using the droop law \cite{ref:2}:
\begin{equation}\label{droop}
	v_n=x_n-r_{n}i_n, \; n=1,...,N,
\end{equation}
where $i_n$ is the output current, and $x_n$ and $r_{n}$ are the \textit{reference voltage} and the \textit{virtual resistance} (i.e., \textit{droop slope}), respectively, whose values are subject to control.
The reference voltage $x_n$ corresponds to the {{rated} voltage of the MG} \cite{ref:3}, while the virtual resistance $r_{n}$ is set to enable proportional load sharing among the DGs.
The nominal values of these parameters, when not using power talk, are denoted with $x_n^{\texttt{n}}$ and $r_{n}^{\texttt{n}}$, corresponding to the nominal bus voltage $v_n^{\texttt{n}}$.
Each bus also hosts a collection of local loads, modeled by a resistance $r_{n}^{\text{cr}}$, constant current $i_{n}^{\text{cc}}$ and constant power $d_{n}^{\text{cp}}$ components connected in parallel, see Fig.~\ref{BusArch}; the loads change {randomly} through time.
The buses are interconnected through DC distribution lines.
The resistance of the line between buses $n$ and $m$ is denoted with $r_{n,m}$, see Fig.~\ref{BusArch}: by convention, $r_{n,m}=\infty$ if $n=m$ or if buses $n$ and $m$ are not directly connected.
We write:
\begin{align}
r_{n}^{\text{bus}} & = \bigg( \frac{1}{r_{m}^{\text{cr}}} + \sum_{m\in\mathcal{N}}\frac{1}{r_{n,m}} + \frac{1}{r_{n}} \bigg)^{-1},
\end{align}
to denote the equivalent bus-to-the-ground {resistance} of bus $n$, including the resistive component of the load.
All voltages, currents, powers and impedances in DC systems are real numbers \cite{ref:12new}.

The physical behavior of the system is governed by the Ohm's and Kirchoff's laws: applying them for the system shown in Fig.~\ref{BusArch}, we obtain the following current balance equation for bus $n=1,...,N$ in steady state \cite{ref:11new}:
\begin{equation}\label{sum_currents}
\frac{x_n-v_n}{r_{n}}=\frac{v_n}{r_{n}^{\text{cr}}} + i_{n}^{\text{cc}} + \frac{d_{n}^{\text{cp}}}{v_n} + \sum_{m\in\mathcal{N}}\frac{v_n-v_m}{r_{n,m}},
\end{equation}
which yields:
\begin{equation}\label{eq:voltage}
\frac{1}{r_{n}^{\text{bus}}}v_n^2 - \bigg(\frac{x_n}{r_{n}} + \sum_{m\in\mathcal{N}}\frac{v_m}{r_{n,m}} - i_{n}^{\text{cc}}\bigg)v_n + d_{n}^{\text{cp}} = 0.
\end{equation}
The unique, physically viable solution to \eqref{eq:voltage} is the positive root; it can be easily verified that the solution is {real} if the droop parameters satisfy the following condition:
\begin{equation}\label{stability}
	x_n\geq r_{n}\bigg(\sqrt{\frac{4d_{n}^{\text{cp}}}{r_{n}^{\text{bus}}}}-\sum_{m\in\mathcal{N}}\frac{v_m}{r_{n,m}}\bigg), \, n=1,...,N.
\end{equation}
This system of inequalities defines the ranges of allowable values for the droop control parameters $x_n$ and $r_{n}$.

\subsection{Discrete Time Linear Signal Model}\label{sec:DiscreteTimeLinearModel}

We proceed by developing a linearized model for all-to-all full duplex communication scenario, where all VSC units simultaneously transmit and receive data.
We assume that the time is slotted in slots of duration $T_S$ and the units are slot-synchronized.\footnote{The duration of the time slot $T_S$ is set to comply with the control frequency of the primary controller; its value is typically of the order of milliseconds to allow the system to establish steady-state.}
In slot $t$, the reference voltage of VSC $n$ is:
\begin{equation}\label{eq:data_stream}
	x_n(t)=x_n^{\texttt{n}}+\Delta x_n(t),\;n=1,...,N,
\end{equation}
with $\Delta x_n(t)$ being the \textit{input} signal.
Then, the resulting deviation of the bus voltage in slot $t$ can be written as: 
\begin{equation}\label{eq:ch_output}
	v_n(t)=v_n + \Delta v_{n}(t),
\end{equation}
where $\Delta v_n(t)$ is the \textit{output} of the communication channel.
VSC $n$ samples the noisy version of $\Delta v_{n}(t)$ with frequency $f_S$ and uses the average of $N_S=T_Sf_S$ samples over the slot $t$ to obtain the observation{\footnote{More precisely, the bus voltage is sampled after the system reaches a steady state and all transient effects diminish.}}:
\begin{align}\label{eq:y}
	\Delta \tilde{v}_{n}(t)=\Delta v_{n}(t) + z_n(t),
\end{align}
where noise $z_n(t)\sim\mathcal{N}(0,\sigma_z^2)$ can be modeled as an additive Gaussian noise \cite{ref:22,ref:23}.
Finally, we assume that the loads in the system changes randomly with a rate that is {much lower than the signaling rate} $T_S^{-1}$ and that the signaling is done over a single realization of the load values.\footnote{Typically, the average time between consecutive load changes in MG systems is of the order of several seconds or even minutes \cite{ref:2,ref:3,ref:4,ref:5,ref:6}.}

Assume that the reference voltage deviations $\Delta x_n(t)$, $n=1,...,N$, are small:
\begin{equation}\label{lin_assump}
	\frac{|\Delta x_n(t)|}{x_n^{\texttt{n}}} \ll 1\Rightarrow\frac{|\Delta v_{n}(t)|}{v_n} \ll 1.  
\end{equation}
Under this assumption, we use the first-order Taylor approximation of \eqref{eq:voltage} around $x_n^{\texttt{n}}$ to obtain the linear model:
\begin{align}\nonumber
 \Delta v_{n}(t)  & \approx \frac{\partial v_n}{\partial \Delta x_{n}(t)}\Delta x_{n}(t) + \sum_{m\in\mathcal{N}}\frac{\partial v_m}{\partial \Delta v_{n}(t)} \Delta v_{m}(t)\\\label{eq:voltage_lin}
								& = r_{n}^{\text{bus}}\kappa_n\frac{\Delta x_{n}(t)}{r_{n}} + r_{n}^{\text{bus}}\kappa_n\sum_{m\in\mathcal{N}}\frac{\Delta v_m(t)}{r_{n,m}},
\end{align}
where $\kappa_n \geq 1$ is given with:
\begin{align}\label{eq:Dm}
\kappa_n = \frac{1}{2}\Bigg(1+\frac{\frac{x_n^{\texttt{n}}}{r_{n}} + \sum_{m}\frac{v_m}{r_{n,m}} - i_{n}^{\text{cc}}}{\sqrt{(\frac{x_n^{\texttt{n}}}{r_{n}} + \sum_{m}\frac{v_m}{r_{n,m}} - i_{n}^{\text{cc}})^2-\frac{4d_{n}^{\text{cp}}}{r_{n}}}}\Bigg).
\end{align}
We introduce the following notation:
\begin{itemize}
\item input $N\times 1$ vector $\Delta\mathbf{x}(t)=[\Delta x_1(t),...,\Delta x_N(t)]^T$,
\item output $N\times 1$ vector $\Delta\mathbf{v}(t)=[\Delta v_{1}(t),...,\Delta v_{N}(t)]^T$,
\item admittance matrix ${\mathbf{\Psi}}$ of dimension $N\times N$, with entries:
\begin{align}\nonumber
\psi_{n,m} & = \left\{
  \begin{array}{lr}
    \sum_{i\in\mathcal{N}}\frac{1}{r_{n,i}}, &  m=n ,\\
    -\frac{1}{r_{n,m}}, &  n\neq m ,
  \end{array}
\right.
\end{align}
\item modified admittance matrix $\breve{\mathbf{\Psi}}$ in which each diagonal entry is multiplied by $\kappa_n^{-1}$, i.e. $\breve{\psi}_{n,n}=\frac{\psi_{n,n}}{\kappa_n}$,
\item the $N\times N$ matrix $\mathbf{Y}=\text{diag}\left\{r_{n}^{-1}\right\}_{n=1,...,N}$,
\item the $N\times N$ matrix $\mathbf{Y}^{\text{cr}}=\left\{(r_{n}^{\text{cr}})^{-1}\right\}_{n=1,...,N}$,
\item the $N\times N$ matrix $\mathbf{K}=\text{diag}\left\{\kappa_n\right\}_{n=1,...,N}$.
\end{itemize}
Using \eqref{eq:voltage_lin} and the above notation, the linearized input-output relation can be compactly written as follows:
\begin{align}\label{eq:lin_model_vec0}
\Delta \mathbf{v}(t) & \approx (\breve{\mathbf{\Psi}} + \mathbf{K}^{-1}(\mathbf{Y}+\mathbf{Y}^{\text{cr}}))^{-1}\mathbf{Y}\Delta\mathbf{x}(t)\\\label{eq:lin_model_vec1}
										 & = \breve{\mathbf{H}}\Delta\mathbf{x}(t).
\end{align}
We refer to the matrix $\breve{\mathbf{H}}$ as the \emph{channel} matrix of the system.
Finally, we obtain the following linear model for the noisy output observed by VSC $n$:
\begin{align}\label{final_signal_model1}
\Delta \tilde{v}_{n}(t) & =\sum_{m=1}^N \breve{h}_{n,m}\Delta x_m(t) + z_n(t),
\end{align}
where $\breve{h}_{n,m}$ is the entry at position $n,m$ of the channel matrix; it can be shown that $\breve{h}_{n,m} > 0$, $\forall n,m$.
We conclude by noting that the resulting linearized power talk communication channel \eqref{final_signal_model1} is an all-to-all full duplex Gaussian Multiple Access Channel \cite{ref:11}, provided that the channel coefficients $\breve{h}_{n,m}$, $m=1,...,N$, are known.

The channel coefficients $\breve{h}_{n,m}$ determine how strongly the input $\Delta x_m$ influences the output observed by VSC $n$;
$\breve{h}_{n,m}$ is a function of 1) the instantaneous values of all loads, 2) the line impedances and 3) the virtual resistances of the VSCs, which are controllable.
Thus, the values of the channel coefficients can be \textit{modified} through the virtual resistances:
\begin{equation}\label{vr}
	r_{n} = r_{n}^{\texttt{n}} + \Delta r_{n},\;n=1,...,N.
\end{equation}
The optimized values of $r_{n}$, denoted with $r_{n}^{*}$, is fixed during the transmission and, in general case, is different from the nominal value $r_{n}^{\texttt{n}}$, i.e., $\Delta r_{n}\neq 0$.
This phenomenon represents a major difference from standard communication scenarios.
In the rest of the paper we assume that the functional relation $\breve{h}_{n,m}(r_{1},...,r_{N})$ {is \textit{known}} $\forall n,m$.
This assumption implies knowledge of the impedances of the distribution lines and the values of the loads, i.e., knowledge on the matrices $\mathbf{\Psi}$, $\mathbf{Y}^{\text{ca}}$ and $\mathbf{K}$ for given power demand. 
Such knowledge can be available \textit{a priori} (typically, in demand-response scenarios, the load values are available through forecast \cite{ref:11new}), or it can be obtained {through estimation of} $\mathbf{\Psi}$, $\mathbf{Y}^{\text{ca}}$ and $\mathbf{K}$ (this aspect is out of the paper scope). 

In the case when the system does not host non-linear loads, i.e., when $d_{n}^{\text{cp}}=0$, $\forall n$, then $\kappa_n=1, \forall n$ and $\mathbf{K} = \mathbf{I}_N$, i.e., it is equal to the $N\times N$ identity matrix.
In this case the linear model \eqref{eq:lin_model_vec0} is exact and obtains the form:
\begin{align}\label{eq:exact_lin_model_vec0}
\Delta \mathbf{v}(t) = ({\mathbf{\Psi}} + \mathbf{Y}+\mathbf{Y}^{\text{cr}})^{-1}\mathbf{Y}\Delta\mathbf{x}(t) = {\mathbf{H}}\Delta\mathbf{x}(t).
\end{align}

\noindent \textbf{Remark:} From an information-theoretic viewpoint, one can actually send additional information by \emph{modulating} the virtual resistances.
In addition, the channel can be also optimized over the reference voltages.
Particularly, generalizing the above communication scheme, we write:
\begin{align}
x_n(t) & = x_n^{\texttt{n}} + \Delta\overline{x}_n + \Delta x_n(t) = \overline{x}_n + \Delta x_n(t),\\
r_n(t) & = r_n^{\texttt{n}} + \Delta\overline{r}_n + \Delta r_n(t) = \overline{r}_n + \Delta r_n(t).
\end{align}
Assuming that $\Delta x_n(t)$ and $\Delta r_n(t)$ are small, relative to $\overline{x}_n$ and $\overline{r}_n$, respectively, and applying Taylor's expansion we obtain the following linearized model:
\begin{equation}
\Delta\tilde{v}_n(t) = \sum_{m=1}^N\big(\breve{h}_{n,m}\Delta x_m(t) + \breve{\varphi}_{n,m}\Delta r_m(t)\big),
\end{equation}
where $\breve{\varphi}_{n,m}$ represent the gains on the channel used to transmit information through the virtual resistances.
It can be shown that $\breve{h}_{n,m} = \breve{h}_{n,m}(\overline{x}_1,...,\overline{x}_N,\overline{r}_1,...,\overline{r}_N)$ and $\breve{\phi}_{n,m} = \breve{\varphi}_{n,m}(\overline{x}_1,...,\overline{x}_N,\overline{r}_1,...,\overline{r}_N)$, i.e., the channel can be optimized over both the virtual resistances \emph{and} the reference voltages.
In this paper, we treat the special case: 1) $\Delta\overline{x}_n=0$, $\overline{x}_n=x_n^{\texttt{n}}$, i.e., the reference voltages are only used for signaling and, 2) $\Delta r_n(t)=0$, i.e., the virtual resistances are not information-carrying, and act only as state variables over which the linearized channel is optimized (for notational convenience, we omit the overline symbol in $\Delta\overline{r}_n=\Delta{r}_n$).

\subsubsection*{Single Bus System} 
We characterize some of the basic properties of the channel coefficients through a simple case of a single bus systems and note that the same observations hold for multibus systems.
This system is a special case of the system depicted in Fig.~\ref{BusArch}, in which the effects of the distribution lines can be neglected and all units are assumed to be connected to a common point and observing the same voltage.
In this case \eqref{final_signal_model1} transforms into:
\begin{align}\label{singlebus1}
\Delta \tilde{v}_{n}(t) = \sum_{m=1}^N \breve{h}_m \Delta x_m(t) + z_n(t),
\end{align}
where the channel coefficients are $\breve{h}_m=\breve{h} \, r_{m}^{-1}$ and where:
\begin{align}\label{singlebus2}
\breve{h}=r^{\text{bus}}\underbrace{\frac{1}{2}\Bigg(1+\frac{\sum_{n=1}^N\frac{x_n^{\texttt{n}}}{r_{n}} - i^{\text{cc}}}{\sqrt{(\sum_{n=1}^N\frac{x_n^{\texttt{n}}}{r_{n}} - i^{\text{cc}})^2-\frac{4d^{\text{cp}}}{r^{\text{bus}}}}}\Bigg)}_{{\kappa}}.
\end{align}
We observe that if the system does not host non-linear, constant power load, i.e., if $d^{\text{cp}}=0$ , then $\kappa=1$ and the channel coefficients satisfy $h_n=r^{\text{bus}}r_{n}^{-1}<1$ and $\sum_{n=1}^N h_n<1$.
In this case, expression \eqref{singlebus1} reduces to a form used in earlier works on power talk \cite{ref:7,ref:8}. 
In general, when $d^{\text{cp}}\neq 0$, the above observation is not necessarily true and the channel coefficient, depending on the value $d^{\text{cp}}$, can be greater than $1$.
However, the value of $\kappa$, which appears as a result of linearization, in practice is very close to $1$, implying that in most practical cases, $h_n<1$.
In fact, from \eqref{singlebus2}, the assumption $\kappa\approx 1$ is valid as long as the condition:
\begin{equation}
r^{\text{bus}}\bigg(\sum_{n=1}^N\frac{x_n^{\texttt{n}}}{r_{n}} - i^{\text{cc}}\bigg)^2 \gg {4d^{\text{cp}}}
\end{equation}
is satisfied.
The physical meaning of this condition is that the constant power component of the load constitutes negligible part of the total bus load.

\subsection{Input Constraints}\label{sec:CommunicationArchitetures}

\begin{figure}
\centering
\includegraphics[width=\columnwidth] {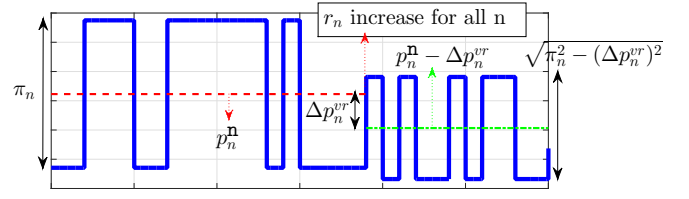}
\caption{{Example of power talk impulses transmitted by unit $n$ in steady state. The full blue line represents the supply power $p_n(t)$ (see eq. \eqref{eq:power_dev1}). The red dash line is the nominal output power $p_n^{\texttt{n}}$ in absence of virtual resistance deviations, i.e., $\Delta p_{n}^{\text{vr}}=0$ and $\Delta p_{n}^{\text{rv}}(t)=0$. Evidently, when the virtual resistance is not deviated, then the power deviation budget $\pi_m$ is completely allocated for the reference voltage power deviations $\Delta p_{n}^{\text{rv}}(t)$. The green dash-dot line is the output power $p_n^{\texttt{n}}+\Delta p_{n}^{\text{vr}}$ after virtual resistance modification and $\Delta p_{n}^{\text{rv}}(t)=0$. In this case, part of the power deviation budget is invested in virtual resistance deviations, and the rest, i.e., $\sqrt{\pi_n^2 - (\Delta p_{n}^{\text{vr}})^2}$, is allocated for power talk signaling. }}
\label{Test}
\end{figure}

Changing the reference voltages and the virtual resistances of VSC units deviates the supplied power.
Thus, we consider imposing constraints on the inputs $\Delta x_n$, $n=1,...,N$, in order to limit the related power deviations.
The output power that VSC $n$ is supplying to bus $n$ is defined as:
\begin{equation}\label{power_def}
p_n=\frac{v_n(x_n-v_n)}{r_{n}}, \; n = 1,...,N.
\end{equation}
During power talk, the output power $p_n(t)$ deviates from the nominal $p_n^{\texttt{n}}$ as follows:
\begin{equation}\label{eq:power_dev1}
p_n(t)=p_n^{\texttt{n}}+\Delta p_n(t), \; n = 1,...,N.
\end{equation}
We bound the average power deviation $\Delta p_n(t)=p_n(t)-p_n^{\texttt{n}}$ with respect to the nominal value $p_n^{\texttt{n}}$:
\begin{align}\label{dev_power_bound1}
\mathbb{E}[(p_n(t)-p_n^{\texttt{n}})^2] & \leq \pi_n^2, \; n = 1,..., N, 
\end{align}
where $\pi_n$ is the \textit{power deviation budget} of VSC $n$.
\begin{figure*}[!b]
\normalsize
\setcounter{MYtempeqncnt}{\value{equation}}
\setcounter{equation}{35}
\hrulefill
\begin{align}\label{new_last}
\breve{\mathbf{H}} = 
\begin{pmatrix}
\begin{bmatrix}
\frac{1}{r_{A,C}} & 0                 & -\frac{1}{r_{A,C}}\\
0                 & \frac{1}{r_{B,C}} & -\frac{1}{r_{B,C}}\\
-\frac{1}{r_{A,C}}&-\frac{1}{r_{B,C}} &  (\frac{1}{r_{A,C}}+\frac{1}{r_{B,C}})\frac{1}{\kappa_C}
\end{bmatrix}
+
\begin{bmatrix}
1 & 0                 & 0\\
0                 & 1 & 0\\
0                 & 0                 & \frac{1}{\kappa_C}
\end{bmatrix}
\begin{pmatrix}
\begin{bmatrix}
\frac{1}{r_{A}} & 0                 & 0\\
0               & \frac{1}{r_{B}}   & 0\\
0               & 0                 & 0
\end{bmatrix}
+
\begin{bmatrix}
0 & 0                 & 0\\
0                 & 0 & 0\\
0                 & 0                 & \frac{1}{r_C^{\text{cr}}}
\end{bmatrix}
\end{pmatrix}
\end{pmatrix}^{-1}
\begin{bmatrix}
\frac{1}{r_{A}} & 0                 & 0\\
0                 & \frac{1}{r_{B}} & 0\\
0                 & 0                 & 0
\end{bmatrix}.
\end{align}
\end{figure*}
\addtocounter{equation}{-5}

The power deviation $\Delta p_n(t)$ can be decomposed as: 
\begin{equation}\label{eq:power_dev22}
\Delta p_n(t) = \Delta p_n^{\text{vr}} + \Delta p_n^{\text{rv}}(t),
\end{equation}
where $\Delta p_n^{\text{vr}}$ is the power deviation due to deviations in the virtual resistances $\Delta r_{n}$, see \eqref{vr}, while $\Delta p_n^{\text{rv}}(t)$ are the power deviations related to the reference voltage deviation $\Delta x_n(t),\;n=1,...,N$, see \eqref{eq:data_stream}, respectively.
In the proposed communication scheme, the virtual resistances are fixed to their optimized values, such that $p_n^{\texttt{n}}+\Delta p_{n}^{\text{vr}}$ is the power supplied by VSC $m$ in absence of the reference voltage deviations.
Afterward, power talk communication through reference voltage deviations is established around the new power supply level.
The key aspect here is that the power deviation bound is defined with respect to the nominal supply level $p_n^{\texttt{n}}$, as formulated in \eqref{dev_power_bound1}.

This is illustrated in Fig.~\ref{Test}.
Evidently, if the channel gains are not modified through the virtual resistances, i.e., when $\Delta r_n = 0$ and $\Delta p_{n}^{\text{vr}}=0$, then the available power budget is allocated only for power talk communication through deviations of the reference voltages.
However, after optimizing the channel gains, the available power budget for communication is reduced, as portion of $\pi_n$ is allocated to deviate the virtual resistances.
This presents a trade-off between the power deviation ``investment'' into the deviations of reference voltages used for communication and in deviations of virtual resistances used for optimizing channel gains. 
Section \ref{sec:CaseStudy} shows how to optimize this trade-off for a simple one-way communication in order to maximize the received SNR.

For small $\Delta x_n(t)$, $n=1,...,N$, the first-order Taylor approximation of $p_n(t)$ around $p_n^{\texttt{n}}+\Delta p_{n}^{\text{vr}}$ is:
\begin{align}\label{power_lin_approx}
\Delta p_{n}^{\text{rv}}(t) 
			       \approx\sum_{m=1}^N\breve{\phi}_{n,m}\Delta x_m(t), \; n=1,...,N,
\end{align}
where:
\begin{align}\label{phi}
\breve{\phi}_{n,m} & =
\left\{
  \begin{array}{lr}
    \frac{\breve{h}_{n,m}x_n^{\texttt{n}}-2\breve{h}_{n,m}v_n}{r_{n}}, &  m\neq n, \\
    \frac{\breve{h}_{n,n}x_n^{\texttt{n}}-2\breve{h}_{n,n}v_n}{r_{n}} + \frac{v_n}{r_{n}}, &  m=n .
  \end{array}
\right.
\end{align}
Assuming that $\mathbb{E}[ \Delta x_n ] = 0$ and $\mathbb{E}[ \Delta x_n \Delta x_m ] = 0$, for $n,m = 1,..., N$, we get:
\begin{align}
\label{dev_power_bound2}
\sum_{m=1}^{N}\phi_{n,m}^2 \, \mathbb{E}[\Delta x_m^2] & \leq\pi_n^2-(\Delta p_{n}^{\text{vr}})^2, \; n = 1,..., N.
\end{align}
Evidently, all communicating units take part in constraining all inputs.
Further, using \eqref{dev_power_bound2} one can formulate a linear program for maximizing $\mathbb{E}[\Delta x_n^2]$, $n=1,...,N$, such that the constraints are met.

We end this section by noting that the developed linear all-to-all full duplex communication model accommodates all possible communication scenarios such as one-way, two-way, broadcast, multicast and multiple access; they can be easily derived from \eqref{eq:lin_model_vec1} by  modifying the respective entries in the input vector $\Delta\mathbf{x}(t)$, depending on whether specific VSCs transmit or not.

\section{Case Study}\label{sec:CaseStudy}

\begin{figure}
\centering
\includegraphics[scale=0.34]{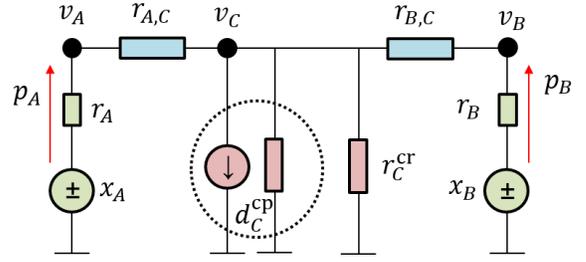}
\caption{Test DC MicroGrid system.}
\label{TestSystem}
\end{figure}

\begin{figure*}[!b]
\normalsize
\setcounter{MYtempeqncnt}{\value{equation}}
\setcounter{equation}{43}
\hrulefill
\begin{equation}\label{eq:bus_vol_ex}
v_n = r_n^{\text{bus}}\begin{pmatrix}\frac{x_n^{\texttt{n}}}{r_n} + \frac{\frac{x_A^{\texttt{n}}}{r_A + r_{A,C}}+\frac{x_B^{\texttt{n}}}{r_B + r_{B,C}} + \sqrt{(\frac{x_A^{\texttt{n}}}{r_A + r_{A,C}}+\frac{x_B^{\texttt{n}}}{r_B + r_{B,C}})^2 - 4d_{C}^{\text{cp}}(\frac{1}{r_A+r_{A,C}} + \frac{1}{r_B+r_{B,C}} + \frac{1}{r_{C}^{\text{cr}}})}}{2r_{n,C}(\frac{1}{r_A+r_{A,C}} + \frac{1}{r_B+r_{B,C}} + \frac{1}{r_{C}^{\text{cr}}})}\end{pmatrix}\;n=A,B.
\end{equation}
\end{figure*}
\addtocounter{equation}{-9}
In this section we focus on the simple case of one-way communication between two VSC units in the system depicted in Fig.~\ref{TestSystem}.
The goal of this example is to illustrate how to take advantage of the property that the channel coefficients can be modified through the virtual resistances. 

The system comprises three buses.
Buses $A$ and $B$ are generation buses that host only VSCs while bus $C$ is a load bus, hosting only remote load.
Bus $C$ is connected to buses $A$ and $B$ through distribution lines.
Without loss of generality, {the load comprises constant resistance and non-linear, constant power component and no constant current load.}
We assume a distance based model for the distribution line resistances, where the resistance of the line is proportional to the line length $L$: $r_{A,C}=\rho \, L_{A,C}$ and $r_{B,C}=\rho \, L_{B,C}$, where {$\rho=0.641\, \Omega/ \text{km}$}~\cite{ref:3,ref:6}, $L_{A,C}=0.3 \, \text{km}$ and $L_{B,C}=1 \, \text{km}$.
The nominal reference voltages of the VSCs are $x_{A}^{\texttt{n}}=x_{B}^{\texttt{n}}=400 \, \text{V}$, the nominal virtual resistances $r_{A}^{\texttt{n}}=r_{B}^{\texttt{n}}=0.39 \,\Omega$.  
We assume that the standard deviation of the observation noise, see \eqref{eq:y}, is $\sigma_z=0.01~ \text{V}$ \cite{ref:10}.


\subsection{Communication Model}\label{sec:OnewayCommunication}

\addtocounter{equation}{1}

Assume that VSC $A$ transmits to VSC $B$, while VSC $B$ is silent, i.e. $\Delta x_B(t) = 0$.
The vectors $\Delta\mathbf{x}(t)$ and $\Delta\mathbf{y}(t)$ in \eqref{eq:lin_model_vec1} can be written as $\Delta\mathbf{x}(t) = [\Delta x_A(t),0,0]^{T}$ and $\Delta\mathbf{v}(t) = [\Delta v_A(t),\Delta v_B(t),\Delta v_C(t)]^{T}$ while the channel matrix $\breve{\mathbf{H}}$ is given by \eqref{new_last} that is displayed {at the bottom of the page}, where:
\begin{equation}
\kappa_C = \frac{1}{2}\Bigg(1+\frac{\frac{v_A}{r_{A,C}} + \frac{v_B}{r_{B,C}}}{\sqrt{(\frac{v_A}{r_{A,C}} + \frac{v_B}{r_{B,C}})^2-\frac{4d_{C}^{\text{cp}}}{r_{C}}}}\Bigg).
\end{equation}
The output observed by VSC B is:
\begin{equation}\label{eq:oneway1}
\Delta \tilde{v}_B(t) = \breve{h}_{B,A}\Delta x_A(t)+z_B(t).
\end{equation}
The VSC B employs maximum likelihood detection on the observed channel output, and the effective received SNR is:
\begin{equation}\label{eq:recSNR1}
\text{SNR}_B=\frac{\breve{h}_{B,A}^2 \, \mathbb{E}[\Delta x_A^2]}{\sigma_z^2}.
\end{equation} 
The constraints \eqref{dev_power_bound2} on input $\Delta x_A$  become:
\begin{align}\label{eq:oneway2}
\breve{\phi}_{A,A}^2 \, \mathbb{E}[\Delta x_A^2] & \leq \pi_A^2-(\Delta p_{A}^{\text{vr}})^2,\\\label{eq:oneway3}
\breve{\phi}_{B,A}^2 \, \mathbb{E}[\Delta x_A^2] & \leq \pi_B^2-(\Delta p_{B}^{\text{vr}})^2,
\end{align}
from which we obtain the following expression for the received SNR:
\begin{align}\label{SNR_gen2}
\text{SNR}_B & = \frac{\breve{h}_{B,A}^2}{{\sigma_z^2}}\min\left\{\frac{\pi_A^2 - (\Delta p_{A}^{\text{vr}})^2}{\breve{\phi}_{A,A}^2},\frac{\pi_B^2 - (\Delta p_{B}^{\text{vr}})^2}{\breve{\phi}_{B,A}^2}\right\}.
\end{align}
The channel gains $\breve{h}_{A,A}$ and $\breve{h}_{B,A}$ can be obtained from $\breve{\mathbf{H}}$ in \eqref{new_last}, after which $\breve{\phi}_{A,A}$ and $\breve{\phi}_{B,A}$ are calculated from \eqref{phi}.
The ``investments'' in power deviations due to virtual resistance modifications are calculated as follows:
\begin{align}
\Delta p_{n}^{\text{vr}} = \frac{(x_n^{\texttt{n}} - v_n)v_n}{r_n} - \frac{(x_n^{\texttt{n}} - v_n^{\texttt{n}})v_n^{\texttt{n}}}{r_n^{\texttt{n}}}, \; n=A,B,
\end{align}
where $v_n$ is given with \eqref{eq:bus_vol_ex}.

\subsection{SNR Maximization}\label{sec:SNRMaximization}
\addtocounter{equation}{1}

We rewrite \eqref{SNR_gen2} as:
\begin{align}\label{SNR_gen3}
\text{SNR}_B & = \frac{1}{{\sigma_z^2}}\min\left\{g_A(r_{A},r_{B}),g_B(r_{A},r_{B})\right\},
\end{align}
where the functions
\begin{align}\label{eq:g_func1}
g_A(r_{A},r_{B})=\frac{\breve{h}_{B,A}^2}{ \breve{\phi}_{A,A}^2} ( \pi_A^2-(\Delta p_{A}^{\text{vr}})^2 ),\\\label{eq:g_func2}
g_B(r_{A},r_{B})=\frac{\breve{h}_{B,A}^2}{ \breve{\phi}_{B,A}^2} ( \pi_B^2-(\Delta p_{B}^{\text{vr}})^2 ),
\end{align}
are introduced for notation convenience.
We are interested in the behavior of the received SNR as function of the virtual resistances over the domain $\mathcal{R}=\left\{(r_{A},r_{B}):g_A\geq 0,g_B\geq 0\right\}$, i.e. for pairs $r_{A},r_{B}$ for which the SNR is positive, which is equivalent to the conditions $\Delta p_{A}^{\text{vr}}<\pi_A$ and $\Delta p_{B}^{\text{vr}}<\pi_B$.
In this respect, it can be shown that:
\begin{itemize}
\item In order to increase the values of the functions $g_A$ and $g_B$, one has to increase $r_A$ and $r_B$ beyond their nominal values.
\item By investigating the Hessian of the vector function $[g_A,g_B]^{T}$ with respect to $[r_A,r_B]^{T}$, it can be verified that these functions are \textit{concave} over $\mathcal{R}$.
\item Finally, the received SNR is the minimum of $g_A$ and $g_B$, and is therefore also a \textit{concave} function over $\mathcal{R}$.
\end{itemize}
Due to the concavity of the SNR, optimal combination $r_{A}^{*},r_{B}^{*}$ that maximizes the received SNR subject to the available power deviation budgets can be found.
The SNR maximization problem can be formally written as:
\begin{align}\label{SNRmaximization}
\max_{r_{A},r_{B}} & g_A,\\\label{const1}
					\text{s.t.}\;\;	 & g_A=g_B,\\\label{const2}
											 & r_{n}^{\texttt{n}}\leq r_{n}\leq r_{n}^{\max}, \; n \in \left\{A,B\right\},
\end{align}
where $r_{n}^{\max}$ is the upper bound of the allowable {dynamic} range on the virtual resistances, satisfying \eqref{stability}.
In other words, we seek the intersection point of the functions $g_A$ and $g_B$ for which \eqref{SNR_gen3} is maximized.

\begin{figure}
\centering
\includegraphics[scale=0.56]{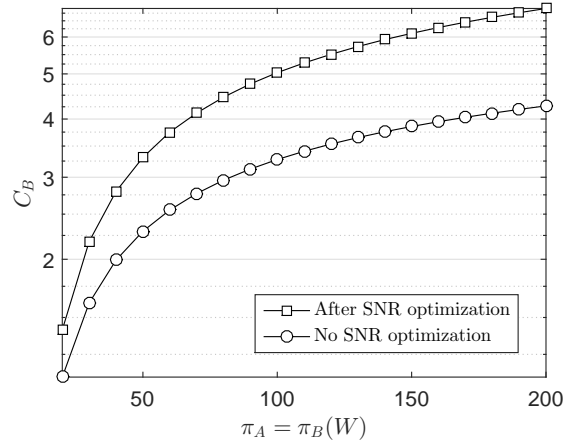}
\caption{The capacity of the scheme in one-way power talk channel \eqref{eq:oneway1} for a given value $\breve{h}_{B,A}$ ($r_{A}^{\texttt{n}}=r_{B}^{\texttt{n}}=0.39 \, \Omega$, $r_{C}^{\text{cr}}=50 \, \Omega$, $d_{C}^{\text{cp}}=2500 \, \text{W}$, $i_{C}^{\text{cc}}=0\,\text{A}$).}
\label{CbvsS}
\end{figure}

The functions $g_A$ and $g_B$ are nonlinear in the virtual resistances, preventing us from finding a closed form solution to \eqref{SNRmaximization}.
Therefore, we resort to iterative, global optimization solver.
In this paper, we employ a grid-search in the region $[r_{A}^{\texttt{n}}\leq r_{A}\leq r_{A}^{\max}, r_{B}^{\texttt{n}}\leq r_{B}\leq r_{B}^{\max}]$, using step of $0.005 \, \Omega$ for both virtual resistances, to obtain $r_{A}^{*},r_{B}^{*}$.
Fig.~\ref{CbvsS} depicts the capacity $C_B$ for given values of the loads $d_{C}^{\text{cp}}$ and $r_{C}^{\text{cr}}$ of the one-way power talk channel \eqref{eq:oneway1} before/after SNR maximization, as a function of the power deviation budget when $\pi_A=\pi_B$:
\begin{equation}\label{capacity}
	C_B=\frac{1}{2}\log(1 + SNR_B).
\end{equation}
Evidently, the capacity of the scheme increases as the available power budget increases.
However, it is also obvious that a significant gain in the information rate can be achieved after optimizing values of the virtual resistances/maximizing the received SNR.

\section{Conclusion}\label{sec:Conclusion}

In this paper, we focused on power talk for multibus DC MG systems of arbitrary configuration, investigating the case in which the information is modulated in the deviations of the reference voltage.
To analyze the proposed communication setup, we developed a small signal framework and showed that the communication channel outputs (i.e., bus voltage deviations) are linear in the input signals (i.e., reference voltages), while the channel coefficients are functions of the virtual resistances.
Through a case study, we showed how to exploit these properties in order to optimize the virtual resistances such that effective SNR is maximized under the constraints on supplied power deviation.

The focus of our future work will be on the general approach in which both controllable parameters, reference voltage $x_n$ and virtual resistance $r_{n}$, are jointly optimized in order to maximize the capacity of the derived communication channels.
The main challenge in this regard is the highly non-linear relation between the channel output and the inputs, consisting of reference voltage and virtual resistance deviations; thus, further investigation is required to develop analytically tractable model.





\section*{Acknowledgment}

The work presented in this paper was supported in part by EU, under grant agreement no. 607774 ``ADVANTAGE''.


\begin{thebibliography}{1}

\bibitem{ref:1}
R.~Lasseter, ``Microgrids,'' in \emph{Power Engineering Society Winter Meeting,
  2002. IEEE}, vol.~1, 2002, pp. 305--308.

\bibitem{ref:2}
J.~Guerrero, J.~Vasquez, J.~Matas, L.~de~Vicuna, and M.~Castilla,
  ``Hierarchical control of droop-controlled {AC} and {DC} microgrids; a general
  approach toward standardization,'' \emph{IEEE Trans. Ind. Electron.}, vol.~58, no.~1, pp. 158--172, Jan. 2011.

\bibitem{ref:3}
T.~Dragicevic; X.~Lu; J.~Vasquez; J.~Guerrero, ``{DC} {M}icrogrids {P}art I: {A Review of Control Strategies and Stabilization Techniques},'' \emph{IEEE Trans. Power Elect.} , vol.~PP, no.~99, pp.1-1. Sep. 2015.

\bibitem{ref:4}
J.~Schonberger, R.~Duke, and S.~Round, ``{DC-bus signaling: A distributed
  control strategy for a hybrid renewable nanogrid},'' \emph{IEEE Trans. Ind. Electron.}, vol.~53, no.~5, pp. 1453--1460, Oct. 2006.

\bibitem{ref:5}
D.~Chen, L.~Xu, and L.~Yao, ``{DC voltage variation based autonomous control of
  DC microgrids},'' \emph{IEEE Trans. Power Del.}, vol.~28, no.~2,
  pp. 637--648, Apr. 2013.
	
\bibitem{ref:6}
T.~Dragicevic, J.~Guerrero, J.~Vasquez, and D.~Skrlec, ``{Supervisory control of
  an adaptive-droop regulated DC microgrid with battery management
  capability},'' \emph{IEEE Trans. Power Electron.}, vol.~29, no.~2,
  pp. 695--706, Feb. 2014.
	
	\bibitem{ref:61}
S.~Galli, A.~Scaglione, and Z.~Wang, ``{For the grid and through the grid: The
  role of power line communications in the smart grid},'' \emph{Proc. IEEE}, vol.~99, no.~6, pp. 998--1027, Jun. 2011.

\bibitem{ref:7}
M.~Angjelichinoski, C.~Stefanovic, P.~Popovski, H.~Liu, P.~Loh, and
  F.~Blaabjerg, ``{Power talk: How to modulate data over a DC micro grid bus
  using power electronics},'' \emph{2015 IEEE Global Communications Conference (GLOBECOM)}, San Diego, CA, 2015, pp. 1-7.
	
\bibitem{ref:8}
M.~Angjelichinoski, C.~Stefanovic, P.~Popovski, and F.~Blaabjerg, ``{Power talk
  in dc micro grids: Constellation design and error probability performance},''
  \emph{2015 IEEE International Conference on Smart Grid Communications (SmartGridComm)}, Miami, FL, 2015, pp. 689-694.
	
\bibitem{ref:hpl}
H. Liu et al., ``Power Talk: A novel power line communication in DC MicroGrid,'' 2016 IEEE 8th International Power Electronics and Motion Control Conference (IPEMC-ECCE Asia), Hefei, 2016, pp. 2870-2874.

\bibitem{ref:9}
M. Angjelichinoski, \v C. Stefanovi\' c, P. Popovski, H. P. Liu, P. C. Loh, and F. Blaabjerg, ``{Multiuser Communication through Power Talk in {DC} MicroGrids},'' \emph{IEEE Journal on Selected Areas in Communications: Special Issue on Power Line Communications and its Integration with the Networking Ecosystem}, vol. 34, no. 7, pp. 2006-2021, July 2016.

\bibitem{ref:mdpi}
M. Angjelichinoski, \v C. Stefanovi\' c, P. Popovski, and F. Blaabjerg, “Communication-Theoretic Model of Power Talk for a Single-Bus DC Microgrid,” Information, vol. 7, no. 1, p. 18, Mar. 2016.

\bibitem{ref:11new}
G.~Giannakis, V.~Kekatos, N.~Gatsis, Seung-Jun Kim, Hao Zhu and B.~Wollenberg, ``{Monitoring and Optimization for Power Grids: A Signal Processing Perspective},'' \emph{IEEE Signal Processing Magazine}, vol.~30, no.~5, pp.107--128, Sept. 2013.

\bibitem{ref:12new}
V. Nasirian, S. Moayedi, A. Davoudi, F.L. Lewis, ``{Distributed Cooperative Control of DC Microgrids},'' \emph{IEEE Transactions on in Power Electronics}, vol.30, no.4, pp.2288--2303, April 2015.

\bibitem{ref:22}
A.~Sangswang and C.~Nwankpa, ``{Random noise in switching DC-DC converter:
  verification and analysis},'' in \emph{Proc. of IEEE ISCAS '03}, Bangkok, Thailand, May 2003.

\bibitem{ref:23}
------, ``Effects of switching-time uncertainties on pulsewidth-modulated power
  converters: modeling and analysis,'' \emph{IEEE Trans. Circuits Syst. I, Fundam. Theory Appl.}, vol.~50, no.~8,
  pp. 1006--1012, Aug. 2003.

\bibitem{ref:11}
A.~El~Gamal and Y.~-H.~Kim, \emph{Network Information Theory.} New York: Cambridge Univ. Press, 2011.

\bibitem{ref:10}
S.~Mazumder, A.~Nayfeh, and D.~Boroyevich, ``Theoretical and experimental
  investigation of the fast- and slow-scale instabilities of a {DC-DC}
  converter,'' \emph{IEEE Trans. Power Electron.}, vol.~16, no.~2,
  pp. 201--216, Mar. 2001.



\end{thebibliography}
\end{document}